# Magnetic anisotropy and valence states in $La_2Co_{1-x}Mn_{1+x}O_6$ (x ≈ 0.23) thin films studied by X-ray absorption spectroscopy techniques

Laura López-Mir[1,*], Regina Galceran[1], Javier Herrero-Martín[2], Bernat Bozzo[1], José Cisneros-Fernández[1], Elisa V. Pannunzio Miner[1], Lluís Balcells[1], Benjamín Martínez[1], and Carlos Frontera[1]

[1]Institut de Ciència de Materials de Barcelona-CSIC, C. dels Til·lers s/n (Campus UAB), E-08193 Cerdanyola del Vallès, Spain
[2]ALBA Synchrotron Light Source, C. de la Llum 2-26, E-08920 Cerdanyola del Vallès, Spain

Abstract

X-ray absorption spectroscopy was used to determine the valence state in $La_2Co_{1-x}Mn_{1+x}O_6$ (x≈0.23) thin films. We found that in spite of the non-stoichiometry, Co is in a divalent state while Mn ions show a mixed valence state. The relation of this finding with the magnetic properties of the films is discussed. X-ray magnetic circular dichroism measurements prove that magnetic anisotropy originates from Co spin-orbit coupling and it is strain-dependent: a strong increase of the angular contribution to the magnetic moment is found when in-plane (out-of-plane) and cell parameters get expanded (compressed). This behavior is reproduced by first order perturbation theory calculations.

## I Introduction

The energy splitting of outermost electron bands in $3d$ transition metal oxides is usually described in terms of crystal field and Hund's coupling. Spin-orbit coupling (SOC) is often overlooked because in most of these compounds it is comparatively weaker than the above mentioned interactions. However, in some cases SOC can play an important role on the macroscopic (magnetic) behavior of these oxides giving rise to a large magnetocrystalline anisotropy. Among $3d$ transition metal oxides, Co is a remarkable well-known example of a strongly anisotropic system [1–3]. Magnetic anisotropy, and especially perpendicular magnetic anisotropy (PMA) in thin films has become very relevant for technological applications like high-density magnetic memories [4]. Within the field of applications, ferromagnetic insulators (FMI) are also gaining attention because they can act as spin polarized sources or spin conductors [5]. Among this type of materials, ferromagnetic (FM) interactions must be of exchange-type (since they cannot be driven by charge carriers); in fact double perovskites combining two different $3d$ metals with $t_{2g}^3 e_g^0$ and $t_{2g}^n e_g^2$ configurations are, according to Goodenough-Kanamori-Anderson rules [6–8], particularly suitable for presenting both ferromagnetism and insulating character. Two examples are $La_2CoMnO_6$ and $La_2NiMnO_6$ insulators [9,10] where magnetic exchange interactions between $Co^{2+}$ or $Ni^{2+}$ ($t_{2g}^5 e_g^2$ and $t_{2g}^6 e_g^2$ respectively) and $Mn^{4+}$ ($t_{2g}^3$) cations are of the FM type.

In the case of $La_2CoMnO_6$ FMI character extends beyond the 1:1 Mn:Co ratio. FM behavior has been found in single crystals [11] and polycrystalline samples [12] of $LaMn_{1-y}Co_yO_6$ with y≈ 0.35. In fact, Barilo et al. [11] reported optimal FM properties in samples with Co content y= 0.36. More importantly, Bull et al. have recently shown, using neutron-diffraction experiments that Co/Mn cationic order for y= 0.35 composition is even better than for y= 0.50 (stoichometric $La_2CoMnO_6$), in the sense that for the former no Co ions are present in the Mn sublattice while in the stoichiometric case about 12% of Mn sublattice sites are occupied by Co. Additionally, magnetic measurements showed a Curie temperature ($T_C$ ≈ 210 K) very similar to that of y= 0.50 samples ($T_C$ ≈ 225 K), and an ordered magnetic moment per cation above $3\mu_B$ [13]. Thus, one can consider that $LaMn_{1-y}Co_yO_3$ system, in a range of y-values from 0.50 down to at least y=0.35, forms a double perovskite structure $La_2Co_{1-x}Mn_{1+x}O_6$ (with x=1-2y) which shows a FMI behavior. A lot of discussion has been addressed on the nature of the magnetic interactions leading to ferromagnetism in $LaCo_yMn_{1-y}O_6$ (y=0.50). However, most of the discrepancies and different interpretations in the literature come from the differences in the actual degree of cationic order in different samples. When cationic ordering exists (in double perovskite structure then FM is explained in terms of superexchange interactions between $Co^{2+}$ and $Mn^{4+}$ as pointed above. However, for y<0.5 (x>0), at least a fraction of Mn ions must reduce towards the trivalent state. The role of this fraction of ions and its effect on magnetic interactions has been scarcely discussed in the literature [11–13].

In a previous work, we reported the existence of a strong PMA in $La_2CoMnO_{6-\varepsilon}$ (LCMO) thin films grown by magnetron sputtering on top of $SrTiO_3$ (STO) as well as a strong dependence of the magnetic anisotropy of these films on their strain state [14]. Our results demonstrated that tensile strain induces a strong PMA while compressive strain drives the easy axis to be in-plane. A dependence of the anisotropy direction on film strain has also been found for different films and in special for Co oxides like $CoFe_2O_4$ and $CoCr_2O_4$ spinel systems [15–19]. More recently, we have determined the exact composition of our thin films by electron probe microanalysis measurements, showing a stoichiometry of the type $La_2Co_{1-x}Mn_{1+x}O_6$ (LCMOx) with x≈0.23. Nonetheless, the magnetic properties of our films agree well with those in the literature for Co deficient LCMO bulk presenting Mn/Co cationic order: They present Curie temperatures around 220 K and a saturation magnetization of $6\mu_B$ /f.u. (based on the double perovskite $La_2Co_{1-x}Mn_{1+x}O_6$ f.u.) [11,12]. Under this stoichiometry, the substitution of Co by Mn opens the question of which is the valence of the substituting Mn and how cationic order is achieved. To shed light into these questions in the first part of this paper, we present X-ray absorption (XAS) spectra at the Mn and Co $L_{2,3}$ edges, known to be very sensitive to the 3d electronic configurations. XAS spectra are very sensitive to valence states: different valences produce clearly differentiated final states in the $2p^63d^m$ to $2p^53d^{m+1}$ absorption process which translate into shifts in the energy position of the absorption peaks of the spectra. In our experiment $L_{2,3}$ absorption edges of Mn and Co atoms are simultaneously recorded by total electron (TEY) and fluorescence yield (TFY). It is well known that TEY is mainly sensitive to the outermost layers while TFY gives information about the bulk of the film.

Meanwhile, the second part of the work focuses on understanding the origin of magnetic anisotropy and its strain dependence. For this purpose, we studied samples grown on STO with different oxygen contents (inducing a change in strain [20]) and on top of $(LaAlO_3)_{0.3}(Sr_2AlTaO_6)_{0.7}$ (LSAT) and $LaAlO_3$ (LAO) substrates using X-ray magnetic circular dichroism (XMCD). This is a unique tool allowing studying separately, in an element-specific way, the orbital and spin contributions to the atomic magnetic moment [19,21]. We show that LCMOx behavior has a magnetocrystalline origin that must be attributed to the combination of the large SOC in $Co^{2+}$ ions with the modification of the crystal field due to the strain.

## II Experimental Details

LCMOx thin films used for XMCD measurements have been prepared using RF-sputtering from a $La_2CoMnO_6$ (stoichiometric) target, further details can be found in Ref. [14]. We have used the very same films than in Ref. [14], in particular samples labeled as B, C, D, E and F in that work (labeled likewise here for the sake of simplicity). In addition, we have studied a thicker LCMO film (70 nm) grown on top STO (labeled here as sample A). Samples B, C, and D were also grown on top of STO substrates, at different conditions and with different post-annealing treatments. Samples A and D have optimized deposition conditions, with the highest oxygen content and highest $T_C$. $CaMnO_3$ (CMO), $LaMnO_3$ (LMO) and $La_{0.7}Sr_{0.3}MnO_3$ (LSMO) films were also grown and used as references for the Mn valence evaluation.

X-ray diffraction measurements performed at KMC-2 beamline (BESSY II synchrotron, Helmholtz Zentrum Berlin, Germany) demonstrated that B, C and D films present a growing tensile strain due to progressive oxygen content [20]. Samples E and F were grown on LSAT and LAO substrates respectively and X-ray diffraction data showed that sample E was fully strained while sample F was partially relaxed [14]. Table 1 summarizes the measured cell parameters and the preparation details of these samples.

XAS and XMCD were measured at the Co and Mn $L_{2,3}$ edges in BL29-BOREAS beamline at ALBA Synchrotron Light Source (Barcelona, Spain) in TEY and TFY modes. The maximum applied magnetic field (parallel to the X-ray beam) was 4 Tesla. Measurements were performed under ultra-high vacuum conditions ($2 \cdot 10^{-10}$ mbar) at different incidence angles. Self-absorption was corrected in data acquired in fluorescence detection mode by applying the procedure described in Ref. [22].

The composition of the samples and the target was measured by electron probe micro-analysis (EPMA) using a CAMECA SX-50 electron microprobe equipped with four wavelength-dispersive X-ray spectrometers (Scientific and Technological Center of the University of Barcelona). Within the error bar of the technique, all samples were found to present a 1:1 La:(Co+Mn) atomic ratio. However, Mn:Co one is clearly above 1:1 in all cases which led us to conclude that these films must be described as

La$_2$Mn$_{1+x}$Co$_{1-x}$O$_6$ with x=0.23(2). EPMA results also showed that the target used had the nominal stoichiometry.

## III Results and Discussion

### III.a *Room temperature XAS*

The Co deficiency found by EPMA implies a charge redistribution in order to fulfill charge neutrality. The most reliable mechanism is the reduction of a part of Mn$^{4+}$ cations towards Mn$^{3+}$, formally expressed as: La$^{3+}$Co$^{2+}_{1-x}$Mn$^{4+}_{1-x}$Mn$^{3+}_{2x}$O$^{2-}_{6}$. Under this hypothesis Mn$^{3+}$ ions would be placed in both Co-sublattice and randomly distributed in the Mn sublattice to compensate for the substitution of Co$^{2+}$. In this scenario, the FMI state displayed by this system cannot be directly understood as driven by superexchange interactions between Co$^{2+}$ and Mn$^{4+}$.

To deepen insight into the actual valence states of Co and Mn in our films we first analyze XAS spectra recorded in TFY mode due to its sensitivity to a far deeper region of the samples [22]. This makes particularly suitable to study sample A (the thickest one, with t≈ 70nm). Figure 1 shows the TFY and TEY spectra together with the LCMO bulk sample spectra (with T$_C$ ≈ 225K) reported in Ref. [23]. TEY and TFY spectra of sample A are very similar between them but also to that of stoichiometric LCMO, for which Co$^{2+}$ in high spin state was settled [23]. From this comparison, we can conclude that the non-stoichiometry introduced in our films does not alter significantly Co valence state: Co ions are in 2+ oxidation state and in high spin (HS) configuration. Moreover, we state that regarding the Co electronic structure, there are no significant differences between surface and bulk of the sample.

Figure 2 shows XAS spectra around Mn $L_{2,3}$ edges in both TFY and TEY detection modes, of LaMnO$_3$, La$_{0.7}$Sr$_{0.3}$MnO$_3$, sample A, and CaMnO$_3$ films. A progressive shift of the Mn $L_3$ main peak with the series of samples can be clearly appreciated in both sets of spectra. One can observe that for sample A this spectral feature lies at an energy value between those in LaMnO$_3$ (642.5 eV) and CaMnO$_3$ (644.1 eV). Thus, following a linear relation between the oxidation state and the position of this absolute maximum the sample A Mn $L_3$ main peak center found at 643.6 eV would point towards a ~3.7 valence. For the case of La$_{0.67}$Sr$_{0.33}$MnO$_3$, whose mixed Mn valence state corresponds to +3.3, Mn $L_3$ maximum lies between those of sample A and LaMnO$_3$ reflecting that the energy shift of this spectral feature is directly proportional to the Mn oxidation state shift from 3+ towards 4+.

Regarding the line shape of Mn $L_{2,3}$ absorption peak of sample A, we found that it is very similar to that published by Burnus et al. [23] for the high-T$_C$ LCMO bulk sample where they establish that their material was formed essentially by Mn$^{4+}$. Nevertheless there are some features that lead us to conclude that our films present a small quantity of Mn$^{3+}$ as the small bump at the Mn $L_3$ pre-edge and the shallower "valley" at ~641.5 eV. More conclusively we could reproduce spectrum of sample A by a linear superposition of LCMO ($T_C \approx 225$) and LMO spectra from ref. [23] with weights of 80% and 20%, respectively. This must not be considered as an absolute quantification of the balance between $Mn^{3+}/Mn^{4+}$, as different correcting (unknown) factors would have to be considered.

It is also interesting to compare TFY and TEY measurements in Fig. 2. In general, the only significant difference between both sets of spectra (especially for sample A and LSMO ones) is a feature appearing at about 640.5 eV. This typically reflects the appearance of a small amount of $Mn^{2+}$ on the film surface [24].

We focus now on the series of samples with different degrees of oxygenation (samples B to F) [20]. Figure 3 shows TEY signal obtained for samples B to F at RT around Co $L_{2,3}$ edges. Except for sample B, the shape of the absorption edge is identical to that found for LCMO-bulk samples [23], and to that found for sample A. Thus, we can conclude that it corresponds to $Co^{2+}$ ion in HS configuration. To examine spectrum of sample B, we followed the same procedure used in Ref. [23] to analyze the spectrum of a bulk sample with poor Co/Mn ordering (low $T_C \approx 150$ K). Figure 4(a) presents the difference spectra resulting from the subtraction of sample D from sample B. A scale factor has been introduced for sample D, enough for making the difference non-negative within the errors. This curve resembles that of $LaCoO_3$ at low temperature and indicates that about 25% of Co ions (in sample B) show a $3d\ t_{2g}^6 e_g^0$ configuration ($Co^{3+}$ in low spin state) [23]. For comparison, the spectra of samples C and D plotted in the inset of Fig. 4(a) appear to be very alike. So, the difference in Co valence state between samples C and D, if any, is clearly much smaller than between samples B and D.

The fraction of trivalent Co ions found in sample B is expected to be compensated by a further reduction of dominating $Mn^{4+}$ ions. Therefore Mn spectra were examined accordingly. Figure 4(b) shows the comparison of Mn $L_{2,3}$ of samples B, C and D. The difference spectrum has its maximum displaced to a lower energy value [indicated by the vertical line in Fig. 4(b)]. In accordance with Ref. [23], this proves that $Mn^{4+}$ in sample B is reduced with respect to sample D.

The origin of these valence changes has been attributed to the presence of antisite disorder in the double perovskite structure [9,25]. This would place a certain amount of Co ions in $Mn^{4+}$ sites. As the size of $Mn^{4+}$ (0.530Å) is considerably smaller than that of $Co^{2+}$ in both HS (0.745Å) or LS (0.65Å), it is forced to move to $Co^{3+}$ in LS (0.545Å). Besides, Mn ions moving to Co sites would have enough space to accommodate one extra electron becoming $Mn^{3+}$ (0.645Å) [26]. This would imply that cationic order in sample B is deficient while in the other samples it is optimal. Here we recall that the main difference in the preparation conditions between sample B and the other samples is the annealing process. This would mean that cationic order improves during the annealing at 900ºC in oxygen atmosphere, in contradiction with studies in bulk which proved that cationic ordering process freezes below 1000ºC [27].

III.b *XMCD results*

Once having established the valences of Co and Mn ions, we now present our analysis of the magnetic properties. Figure 5 shows the x-ray absorption and x-ray magnetic circular dichroism spectra at Co $L_{2,3}$ edges of samples B, D, E and F as measured by TEY at 20 K. All spectra correspond to normal incidence geometry (i.e. with the photons' propagation vector parallel to the vector defining the sample surface). The integral curves of the XMCD spectra are also plotted. In spite of the difference in Co valence between samples B and D, their XMCD signals are spectrally very similar. This fact can be easily understood as the extra $Co^{3+}$ detected in sample B is in $t_{2g}^6 e_g^0$ non-magnetic low spin state.

From the sum rules [21,28] we derive the ratio:

$$\frac{m_L}{m_{Seff}} = \frac{L_z}{2S_z + 7T_z} = \frac{-2\int_{L_3+L_2}(\mu_+ - \mu_-)}{3\left[\int_{L_3}(\mu_+ - \mu_-) - 2\int_{L_2}(\mu_+ - \mu_-)\right]} \tag{1}$$

where $L_z$ and $S_z$ denote the projections of angular and spin magnetic moment over the magnetic field direction. $T_z$ is the magnetic dipole moment that has been estimated for $Co^{2+}$ in octahedral environment to be between 2 and 3 orders of magnitude smaller than $S_z$ [29]. The values of the ratio $m_L/m_{Seff}$ hardly depend on the point where the end of $L_3$ and the start of $L_2$ edges are taken. The uncertainty that this introduces to this value has been checked to be below 1%. This value is much smaller than error introduced by other sources [30].

The $m_L/m_{Seff}$ ratios derived from the analysis of the XMCD curves of Co edge obtained on the different samples at two incidence angles (namely normal and at 20º, hereafter referred to as grazing incidence) are listed in Table I. The values obtained at normal incidence present a monotonous behavior with the strain. With the exception of sample F (on top of LAO), the enlargement (shrink) of in-plane (out-of-plane) lattice parameter is accompanied with an increase of $m_L/m_{Seff}$ ratio. On the contrary, at grazing incidence this ratio does not show any clear tendency and its dependence on the strain is much smaller. As a consequence, for compressive strain the $m_L/m_{Seff}$ ratio found in grazing incidence is larger than that found at normal incidence, while for tensile strain this is reversed.

The $m_L/m_{Seff}$ ratios derived from analysis of Mn-$L_{3,2}$ edge are two orders of magnitude smaller than those of Co. So, we can conclude that Mn does not significantly contribute to magnetic anisotropy of our films.

# IV Spin-orbit coupling and crystal field as first order perturbation theory

In order to explain the changes in $m_L/m_{Seff}$ ratio of Co for the different samples, we reproduce the model introduced in Ref. [14]. This model starts from the case where $CoO_6$ octahedra have a perfect cubic symmetry and introduces the tetragonal distortion and spin orbit coupling (SOC) as a perturbation in a procedure similar to that used for

CoCl$_2$ [31,32]. In CoCl$_2$ case, octahedra lose cubic symmetry due to a trigonal distortion induced by a compression/expansion along one of the main diagonals of the cube.

A cubic crystal field produced by a perfect octahedral environment, splits the ground state of a free Co$^{2+}$ ($^4F$ term with L=3, S=3/2) in three levels, 2 triplets and one singlet. The lowest level corresponds to the triplet $^4T_1$ whose eigenstates are [33]:

$$\varphi_0 = |30\rangle$$
$$\varphi_+ = \sqrt{3/8}\,|3\,\text{-}1\rangle + \sqrt{5/8}\,|3\,3\rangle \qquad (2)$$
$$\varphi_- = \sqrt{3/8}\,|3\,1\rangle + \sqrt{5/8}\,|3\,\text{-}3\rangle$$

We now introduce the tetragonal distortion of the octahedron ($H_{CF}$) and the SOC interaction ($H_{LS}$) as perturbations $H' = H_{LS} + H_{CF}$ of the Hamiltonian. In order to apply first order perturbation theory, we calculate the matrix elements of $H'$ in the ground state of unperturbed Hamiltonian. SOC term, $H_{LS}$, is expressed as $k\lambda \vec{L}\cdot\vec{S}$, where $\lambda$ is the spin-orbit constant, that is expected to be negative (for more than half filled $d$-shells), and $k$ is the "orbital reduction factor" ($k \lesssim 1$) [31,32]. $H_{CF}$ is the deviation crystal field from cubic symmetry $\left(H_{CF}^{tet} - H_{CF}^{cub}\right)$. According to Ref. [31] a general tetragonal crystal field is expressed as:

$$H_{CF}^{tet} = A_2^0 r^2 Y_2^0 + A_4^0 r^4 Y_4^0 + r^4\left(A_4^4 Y_4^4 + (A_4^4)^* Y_4^{-4}\right)$$

while for a cubic symmetry it reduces to $H_{CF}^{cub} = A_4^0 r^4\left[Y_4^0 + \left(\frac{5}{14}\right)^{1/2}\left(Y_4^4 + Y_4^{-4}\right)\right]$. Thus, the tetragonal field differs from the cubic one in a term on $Y_2^0$ and on the fact that of $A_4^0$ and $A_4^4$ are no longer related. We have, in a first approximation, ignored this second fact and considered only the first one. Using this approximation, to calculate $H_{ij}^{CF} = \langle\varphi_i|H_{CF}|\varphi_j\rangle$ matrix elements one must calculate $\langle\varphi_i|A_2^0 r^2 Y_2^0|\varphi_j\rangle$. The radial part is the same for all the matrix elements as it does not depend on $m$ but only on $n$ and $l$ quantum numbers. Moreover, it is different from zero, as the integral will only contain positive terms. Concerning the angular part, by conservation of the third component of the angular moment the unique terms that can be different from zero are those coming from: $\langle 3\,0|Y_2^0|3\,0\rangle$, $\langle 31|Y_2^0|31\rangle$, $\langle 3\text{-}1|Y_2^0|3\text{-}1\rangle$, $\langle 33|Y_2^0|33\rangle$, and $\langle 3\text{-}3|Y_2^0|3\text{-}3\rangle$. Their angular parts are given by the integrals $\int_\Omega Y_3^{-m} Y_2^0 Y_3^m d\Omega$ that can be calculated trough the 3-j symbols:

$$\int_\Omega Y_{j_1}^{m_1} Y_{j_2}^{m_2} Y_{j_3}^{m_3} d\Omega = \sqrt{\frac{(2j_1+1)(2j_2+1)(2j_3+1)}{4\pi}}\begin{pmatrix}j_1 & j_2 & j_3\\0 & 0 & 0\end{pmatrix}\begin{pmatrix}j_1 & j_2 & j_3\\m_1 & m_2 & m_3\end{pmatrix}$$

(3)

Taking all this into account, the unique matrix elements of $H_{CF}$ different from zero are: $H_{00}^{CF} = 2\epsilon_{CF}$ and $H_{++}^{CF} = H_{--}^{CF} = -\epsilon_{CF}$. This means that (before considering SOC), $\varphi_0$ state, or equivalently $|30\rangle$, that is mainly oriented along z axis, is more affected by the tetragonal distortion of the crystal field than $\varphi_{\pm 1}$. In the case of tensile strain, where basal distances of the octahedra are larger than apical ones, $\epsilon_{CF}$ is positive and $|30\rangle$ is the state with higher energy while the other two states of the $^4T_1$ triplet have a lower energy level. This is in agreement with the expected degeneration of the ground state under tensile stress [15]. In the case of compressive strain $\epsilon_{CF}$ is negative, and $|30\rangle$ becomes the ground state (crystal field only).

To consider the spin orbit interaction, $H_{ij}^{LS}$ matrix elements must be calculated by using:

$\vec{L} \cdot \vec{S} = L_z S_z + \frac{1}{2}(L_+ S_- + L_- S_+)$ and $L_\pm |LM\rangle = \sqrt{(L \pm M + 1)(L \mp M)}|L \pm 1\rangle$ (and the equivalent for $S_\pm$ operators).

The three levels of $^4T_1$ term must be combined with the four possible spin states (S=3/2) thus giving rise to twelve states. The obtained matrix is given in Table II. The diagonalization of this matrix renders that the lowest energy level is a Kramers doublet that corresponds to (assuming $\lambda<0$):

$$\psi_- = \alpha|\varphi_- -\tfrac{3}{2}\rangle + \beta|\varphi_0 -\tfrac{1}{2}\rangle + \gamma|\varphi_+ \tfrac{1}{2}\rangle; \quad \psi_+ = \alpha|\varphi_+ \tfrac{3}{2}\rangle + \beta|\varphi_0 \tfrac{1}{2}\rangle + \gamma|\varphi_- -\tfrac{1}{2}\rangle \tag{4}$$

being $\alpha$, $\beta$, and $\gamma$ coefficients that only depend on $a = \frac{2\epsilon_{CF}}{-3k\lambda}$:

$$\alpha = \frac{1}{N}\left[-\frac{2}{\sqrt{3}} + \frac{1}{2\sqrt{3}}(1 + 2a + 2\Xi)(2a - \Xi)\right]$$

$$\beta = -\frac{1}{N}\frac{1}{2\sqrt{2}}(1 + 2a + 2\Xi)$$

$$\gamma = \frac{1}{N}$$

(5)

With $N$ being the appropriate normalization factor and $\Xi$ the smaller real solution of the equation:

$$-15 - 20a - 16a^2 - 8a^3 + \Xi(-11 - 8a + 12a^2) + 8\Xi^2 + 4\Xi^3 = 0 \tag{6}$$

XMCD measurements are done under the application of a magnetic field which splits the doublet. This is usually considered as a second perturbation and first order perturbation theory is applied within the Kramers doublet subspace [32]. For simplicity, we consider that the field is applied in an arbitrary direction of the x-z plane $[\vec{H} = H(\cos\theta, 0, \sin\theta)]$. It can be shown that the inclusion of a $y$ component $[\vec{H} = H(\cos\theta\cos\varphi, \cos\theta\sin\varphi, \sin\theta)]$ does not alter the result. The Zeeman Hamiltonian is expressed as:

$$H_Z = \mu_B(k\vec{L} + 2\vec{S}) \cdot \vec{H} = \mu_B H\{(kL_z + 2S_z)\sin\theta + [k\tfrac{1}{2}(L_+ + L_-) + (S_+ + S_-)]\cos\theta\} \tag{7}$$

Matrix elements within Karmers doublet subspace are:

$$H_Z = \begin{pmatrix} -\zeta_1 & \zeta_2 \\ \zeta_2 & \zeta_1 \end{pmatrix}, \quad (8)$$

where

$$\zeta_1 = \left[\left(3 + \frac{3}{2}k\right)\alpha^2 + \beta^2 - \left(1 + \frac{3}{2}k\right)\gamma^2\right]\sin\theta$$

$$\zeta_2 = \left[2\alpha\gamma\sqrt{3} + k\beta\gamma\frac{3}{2}\sqrt{2} + 2\beta^2\right]\cos\theta$$

(9)

Thus rendering that the ground state is a combination of the two states of the Kramers doublet:

$$\xi = \frac{1}{N}(\psi_- + \Delta\psi_+) \text{ with } \Delta = \frac{\zeta_2}{\zeta_1 + \sqrt{\zeta_1^2 + \zeta_2^2}} \text{ and } N = \sqrt{1 + \Delta^2}.$$

In order to obtain the predicted value of $m_L/m_S$ we need to calculate the expected value of the projections of $\vec{L}$ and $\vec{S}$ over the direction of light propagation (parallel to the applied field):

$$\frac{m_L}{m_S} = \frac{\langle L_x \cos\theta + L_z \sin\theta \rangle}{\langle S_x \cos\theta + S_z \sin\theta \rangle}$$

$$= \frac{\Delta 3\gamma\beta\sqrt{2}\cos\theta + (\Delta^2 - 1)\frac{3}{2}(\alpha^2 - \gamma^2)\sin\theta}{\Delta(2\alpha\gamma\sqrt{3} + 2\beta^2)\cos\theta + (\Delta^2 - 1)\left(\frac{3}{2}\alpha^2 + \frac{1}{2}\beta^2 - \frac{1}{2}\gamma^2\right)\sin\theta}$$

(10)

All coefficients in this expression only depend on a single free parameter: $a = \frac{2\epsilon_{CF}}{-3k\lambda}$. This parameter contains the energy of the tetragonal crystal field (the departure of the crystal field from the cubic symmetry) and the spin-orbit coupling coefficient.

In order to compare with experimental data we need to make an estimation of $\epsilon_{CF}$ or, at least, how it varies with structural parameters. We recall that $\epsilon_{CF}$ parametrizes the deviation of Co octahedra from a perfect cubic environment. As far as the film cell is tetragonal when films are fully strained to cubic substrates (STO and LSAT cases [14,20]), we can, at least for low values of the distortion, consider it to be proportional to $(a_F-c_F)/a_F$ (where $a_F$ and $c_F$ are the in-plane and out of plane cell parameters of the film, respectively). The idea behind this approximation is that for low values of the distortion its main effect is to contract/expand Co-O bond distances rather than inducing a bond bending. As the film cell becomes tetragonal when films are fully strained to cubic substrates, this parameter is described by a term rendering the tetragonal distortion of the cell. In other words, we assume that octahedra are cubic ($\epsilon_{CF} = 0$) when $c_F=a_F$, while a tensile stress makes $c_F<a_F$ and $\epsilon_{CF} > 0$ and a compressive strain makes $c_F>a_F$ and $\epsilon_{CF} < 0$. Figure 6 plots the $m_L/m_{Seff}$ values found by XMCD as a function of $(a_F-c_F)/a_F$ as calculated from the cell parameters obtained by X-ray diffraction (top x-axis) and compares them to the values of $m_L/m_S$ calculated from the previous expression as a function of $\frac{2\epsilon_{CF}}{-3k\lambda}$ (bottom x-axis). For these calculations, we chose $k$=0.94.

# V Summary and Conclusions

To summarize, we have presented a study of the local valence and magnetic anisotropy properties of epitaxial films of LCMO by using x-ray spectroscopy techniques. We confirmed from XAS measurements that in LCMO films with a high Curie temperature ($T_C$≈225K), the valence state of Co is essentially 2+, independently of their strain state. Despite electron probe microanalysis shows a deficiency of Co from its nominal compound La$_2$Co$_{1-x}$Mn$_{1+x}$O$_6$ (with x ≈0.23), this does not seem to affect the valence of Co ions, but induces a reduction of Mn oxidation state from 4+ to 3+ to fulfil charge neutrality thus, becoming formally La$^{3+}$Co$^{2+}_{1-x}$Mn$^{4+}_{1-x}$Mn$^{3+}_{2x}$O$^{2-}_6$. In this scenario, divalent Co, high $T_C$, and high saturation magnetization (in optimized films) rule out a disordered arrangement of Co and Mn ions in the double perovskite structure, which reinforce that FM is induced by superexchange interactions.

Some more light on the valence state of Co and Mn can be shed by considering bond distances found by Bull et al. [13] for La$_2$Co$_{0.7}$Mn$_{1.3}$O$_6$ (LaCo$_{0.35}$Mn$_{0.65}$O$_3$) the Wyckoff positions (WP) of $P\,2_1/n$ space group $2c$ (occupied by Co and Mn in a 0.7:0.3 ratio, with ⟨d$_{2c\text{-}O}$⟩=2.027Å) and $2d$ (occupied by Mn only, with ⟨d$_{2d\text{-}O}$⟩=1.923Å). Assuming that $2c$ and $2d$ WP are respectively occupied by Co$^{2+}_{0.7}$Mn$^{3+}_{0.3}$ and Mn$^{4+}_{0.7}$Mn$^{3+}_{0.3}$ respectively, and taking into account bond distances reported for Co$^{2+}$ [34], Mn$^{3+}$ [35] and Mn$^{4+}$ [36], one would expect ⟨d$_{2c\text{-}O}$⟩=2.096Å and ⟨d$_{2d\text{-}O}$⟩=1.935Å. The fact that both experimental bond lengths are slightly smaller than expected could signal that a small fraction of Co$^{2+}$ oxidizes to Co$^{3+}$ in low spin. Such fraction would be below the detection limit of XAS (samples C and D with high $T_C$).

We show that the only film presenting a small $T_C$ (≈150K), due to low oxygen content, has a significant amount (≈25%) of trivalent Co ions in low spin state. The origin of this valence change has been attributed to the presence of antisite disorder in the double perovskite structure.

On the other hand, the XMCD signal for the different samples in either normal or grazing incidence conditions, evidence a large contribution from the orbital angular moment of Co ions, as expected for Co$^{2+}$, but no contribution from Mn ions. Moreover, our data show a strong dependence of the $m_L/m_{Seff}$ ratio on the film strain, notably for the normal incidence case. This is less marked when probing Co 3$d$ empty states with a large out-of-plane component symmetry. The comparison of the $m_L/m_S$ ratio theoretically calculated and XMCD-derived $m_L/m_{Seff}$ values is remarkably good for the normal incidence case. Nevertheless, several features are also qualitatively well reproduced by the predicted curve for the grazing incidence case. First, theory predicts a much smaller dependence on strain for the latter case than when probing in-plane orbitals. Second, theory predicts the intersection of the normal and grazing $m_L/m_S$ curves when inverting the sign of the crystal field term: it predicts that $m_L/m_S$ at normal incidence is smaller (larger) than $m_L/m_S$ at grazing incidence for compressive (tensile) strain.

So, in conclusion, anisotropy phenomena in LCMO films are mainly driven by spin orbit coupling of $Co^{2+}$ in HS state. Most of the features presented can be well explained by starting from $Co^{2+}$ in a perfect octahedral local environment and adding the effect of spin orbit coupling and a small tetragonal distortion from the cubic crystal field as perturbations.

## Acknowledgements

We acknowledge financial support from the Spanish Ministry of Economy and Competitiveness through the "Severo Ochoa" Programme for Centres of Excellence in R&D (SEV-2015-0496), and projects MAT2012-33207 and MAT2015-71664-R. These experiments were performed in the BOREAS beamline at the ALBA Synchrotron Light Source with the collaboration of ALBA staff. L.L.-M. work has been done as a part of her Ph.D. program in Physics at Universitat Autònoma de Barcelona. We thank X. Llobet from the Scientific and Technological Center of the University of Barcelona for the EPMA analysis.

*Contact author: llopez@icmab.es

## Bibliography

[1] R. F. Pearson, Proc. Phys. Soc. 505 (1959).
[2] J. C. Slonczewski, Phys. Rev. 1341 (1958).
[3] H. Shenker, Phys. Rev. 1246 (1957).
[4] S. Mangin, D. Ravelosona, J. A. Katine, M. J. Carey, B. D. Terris, and E. E. Fullerton, Nat. Mater. **5**, 210 (2006).
[5] Y. Kajiwara, K. Harii, S. Takahashi, J. Ohe, K. Uchida, M. Mizuguchi, H. Umezawa, H. Kawai, K. Ando, K. Takanashi, S. Maekawa, and E. Saitoh, Nature 262 (2010).
[6] J. B. Goodenough, Phys. Rev. **100**, 564 (1955).
[7] J. Kanamori, J. Phys. Chem. Solids **10**, 87 (1959).
[8] P. W. Anderson, Phys. Rev. **79**, 350 (1950).
[9] R. I. Dass and J. B. Goodenough, Phys. Rev. B 14401 (2003).
[10] R. I. Dass, J.-Q. Yan, and J. B. Goodenough, Phys. Rev. B **68**, 64415 (2003).
[11] S. N. Barilo, V. I. Gatalskaya, S. V. Shiryaev, L. A. Kurochkin, S. N. Ustinovich, H. Szymczak, R. Szymczak, and M. Baran, Phys. Status Solidi **199**, 484 (2003).
[12] I. O. Troyanchuk, L. S. Lobanovsky, D. D. Khalyavin, S. N. Pastushonok, and H. Szymczak, J. Magn. Magn. Mater. **210**, 63 (2000).
[13] C. L. Bull, H. Y. Playford, K. S. Knight, G. B. G. Stenning, and M. G. Tucker, Phys. Rev. B **94**, 14102 (2016).
[14] R. Galceran, L. López-Mir, B. Bozzo, J. Cisneros-Fernández, J. Santiso, L. Balcells, C. Frontera, and B. Martínez, Phys. Rev. B **93**, 144417 (2016).

[15] J. A. Heuver, A. Scaramucci, Y. Blickenstorfer, S. Matzen, N. A. Spaldin, C. Ederer, and B. Noheda, Phys. Rev. B 214429 (2015).
[16] P. C. Dorsey, P. Lubitz, D. B. Chrisey, and J. S. Horwitz, J. Appl. Phys. 6338 (1996).
[17] A. Lisfi, C. M. Williams, L. T. Nguyen, J. C. Lodder, A. Coleman, H. Corcoran, A. Johnson, P. Chang, A. Kumar, and W. Morgan, Phys. Rev. B 54405 (2007).
[18] G. Hu, J. H. Choi, C. B. Eom, V. G. Harris, and Y. Suzuki, Phys. Rev. B (2000).
[19] C. Gatel, B. Warot-Fonrose, S. Matzen, and J.-B. Moussy, Appl. Phys. Lett. **103**, 92405 (2013).
[20] R. Galceran, C. Frontera, L. Balcells, J. Cisneros-Fernández, L. López-Mir, J. Roqueta, J. Santiso, N. Bagués, B. Bozzo, A. Pomar, F. Sandiumenge, and B. Martínez, Appl. Phys. Lett. **105**, 242401 (2014).
[21] P. Carra, B. T. Thole, M. Altarelli, and X. Wang, Phys. Rev. Lett. 694 (1993).
[22] S. Eisebitt, T. Böske, J.-E. Rubensson, and W. Eberhardt, Phys. Rev. B **47**, 14103 (1993).
[23] T. Burnus, Z. Hu, H. H. Hsieh, V. L. J. Joly, P. A. Joy, M. W. Haverkort, H. Wu, A. Tanaka, H.-J. Lin, C. T. Chen, and L. H. Tjeng, Phys. Rev. B **77**, 125124 (2008).
[24] S. Valencia, A. Gaupp, W. Gudat, L. Abad, L. Balcells, A. Cavallaro, B. Martínez, and F. J. Palomares, Phys. Rev. B **73**, 104402 (2006).
[25] T. Kyômen, R. Yamazaki, and M. Itoh, Chem. Mater. (2003).
[26] R. D. Shannon, Acta Cryst. **A32**, 751 (1976).
[27] T. Kyômen, R. Yamazaki, and M. Itoh, Chem. Mater. (2004).
[28] B. T. Thole, P. Carra, F. Sette, and G. van der Laan, Phys. Rev. Lett. 1943 (1992).
[29] Y. Teramura, A. Tanaka, B. Thole, and T. Jo, J. Phys. Soc. Japan **65**, 3056 (1996).
[30] C. Piamonteze, P. Miedema, and F. M. F. de Groot, Phys. Rev. B 184410 (2009).
[31] M. E. Lines, Phys. Rev. 546 (1963).
[32] D. Dai, H. Xiang, and M.-H. Whangbo, J. Comput. Chem. 2187 (2008).
[33] B. Bleaxey and K. W. H. Stevens, Rep. Prog. Phys. 108 (1953).
[34] W. Jauch, M. Reehuis, H. J. Bleif, F. Kubanek, and P. Pattison, Phys. Rev. B **64**, 52102 (2001).
[35] C. Ritter, M. R. Ibarra, J. M. De Teresa, P. A. Algarabel, C. Marquina, J. Blasco, J. García, S. Oseroff, and S.-W. Cheong, Phys. Rev. B **56**, 8902 (1997).
[36] E. S. Božin, A. Sartbaeva, H. Zheng, S. A. Wells, J. F. Mitchell, T. Proffen, M. F. Thorpe, and S. J. L. Billinge, J.~Phys.~Chem.~of Solids **69**, 2146 (2008).

Table I: Description of the different samples used for the study: substrate, annealing conditions (oxygen pressure and time), together with the out-of-plane lattice parameter obtained by X-ray diffraction [14,20]. All samples were grown at a partial oxygen pressure of 0.4 Torr and annealed at 900ºC except sample B (grown at 0.3 Torr and not annealed). The two last columns on the right contain $m_L/m_{Seff}$ obtained by XMCD at $T=$ 20K at normal incidence and 20º incidence.

| Name | Subs. | Thickness (nm) | Ann. $p_{O2}$ (Torr) | Ann. time (h) | Cooling rate (ºC/min) | Lat. Par. (Å) | $m_L/m_{Seff}$ (90º) | $m_L/m_{Seff}$ (20º) |
|------|-------|----------------|----------------------|---------------|-----------------------|---------------|----------------------|----------------------|
| A | STO | 70 | 400 | 2 | 10 | | | |
| B | STO | 15 | - | 0 | 10 | 3.902(3) | 0.567 | 0.581 |
| C | STO | 15 | 400 | 1 | 10 | 3.881(3) | 0.608 | - |
| D | STO | 15 | 400 | 1 | 1 | 3.868(3) | 0.637 | 0.560 |
| E | LSAT | 15 | 400 | 2 | 10 | 3.906(5) | 0.456 | 0.558 |
| F | LAO | 15 | 400 | 2 | 10 | 3.910(5) | 0.485 | 0.432 |

Table II: Matrix elements of the perturbation Hamiltonian $H'=H_{LS}+H_{CF}$ (divided by a common factor $\frac{-3k\lambda}{2}$).

| | $|\varphi_- -3/2\rangle$ | $|\varphi_- -1/2\rangle$ | $|\varphi_- 1/2\rangle$ | $|\varphi_- 3/2\rangle$ | $|\varphi_0 -3/2\rangle$ | $|\varphi_0 -1/2\rangle$ | $|\varphi_0 1/2\rangle$ | $|\varphi_0 3/2\rangle$ | $|\varphi_+ -3/2\rangle$ | $|\varphi_+ -1/2\rangle$ | $|\varphi_+ 1/2\rangle$ | $|\varphi_+ 3/2\rangle$ |
|---|---|---|---|---|---|---|---|---|---|---|---|---|
| $\langle\varphi_- -3/2|$ | -a-3/2 | 0 | 0 | 0 | 0 | -√(3/2) | 0 | 0 | 0 | 0 | 0 | 0 |
| $\langle\varphi_- -1/2|$ | 0 | -a-1/2 | 0 | 0 | 0 | 0 | -√2 | 0 | 0 | 0 | 0 | 0 |
| $\langle\varphi_- 1/2|$ | 0 | 0 | -a+1/2 | 0 | 0 | 0 | 0 | -√(3/2) | 0 | 0 | 0 | 0 |
| $\langle\varphi_- 3/2|$ | 0 | 0 | 0 | -a+3/2 | 0 | 0 | 0 | 0 | 0 | 0 | 0 | 0 |
| $\langle\varphi_0 -3/2|$ | 0 | 0 | 0 | 0 | 2a | 0 | 0 | 0 | 0 | -√(3/2) | 0 | 0 |
| $\langle\varphi_0 -1/2|$ | -√(3/2) | 0 | 0 | 0 | 0 | 2a | 0 | 0 | 0 | 0 | -√2 | 0 |
| $\langle\varphi_0 1/2|$ | 0 | -√2 | 0 | 0 | 0 | 0 | 2a | 0 | 0 | 0 | 0 | -√(3/2) |
| $\langle\varphi_0 3/2|$ | 0 | 0 | -√(3/2) | 0 | 0 | 0 | 0 | 2a | 0 | 0 | 0 | 0 |
| $\langle\varphi_+ -3/2|$ | 0 | 0 | 0 | 0 | 0 | 0 | 0 | 0 | -a+3/2 | 0 | 0 | 0 |
| $\langle\varphi_+ -1/2|$ | 0 | 0 | 0 | 0 | -√(3/2) | 0 | 0 | 0 | 0 | -a+1/2 | 0 | 0 |
| $\langle\varphi_+ 1/2|$ | 0 | 0 | 0 | 0 | 0 | -√2 | 0 | 0 | 0 | 0 | -a-1/2 | 0 |
| $\langle\varphi_+ 3/2|$ | 0 | 0 | 0 | 0 | 0 | 0 | -√(3/2) | 0 | 0 | 0 | 0 | -a-3/2 |

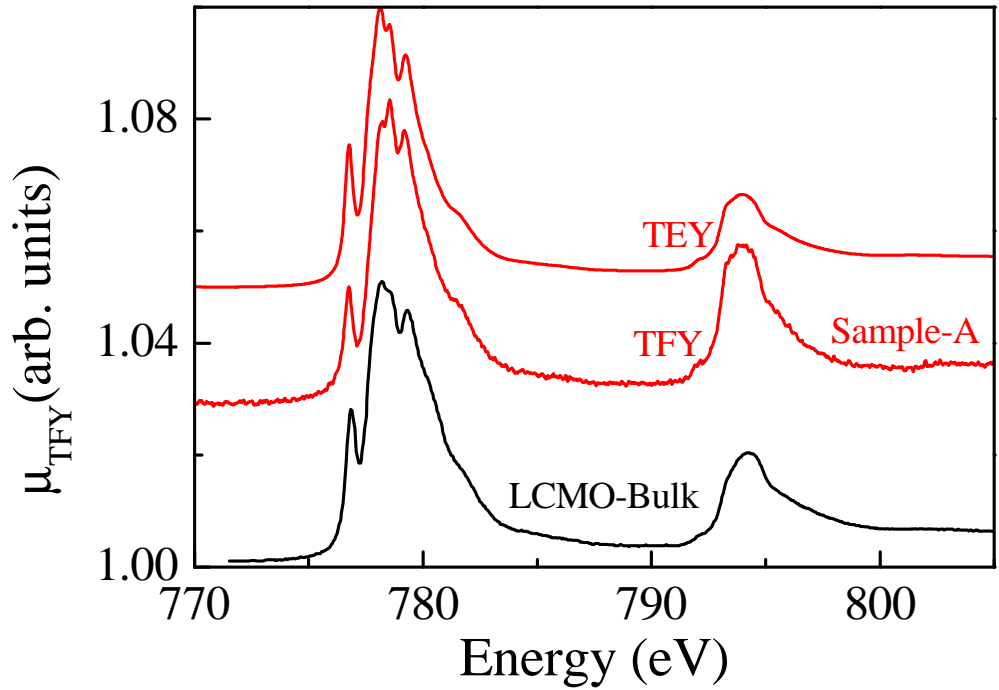

Figure 1 Co $L_{2,3}$ XAS of Sample A (red lines, comparing TFY and TEY signals) and LCMO bulk (as extracted from Ref. [23], black line) at 300 K. Spectra have been vertically shifted for clarity.

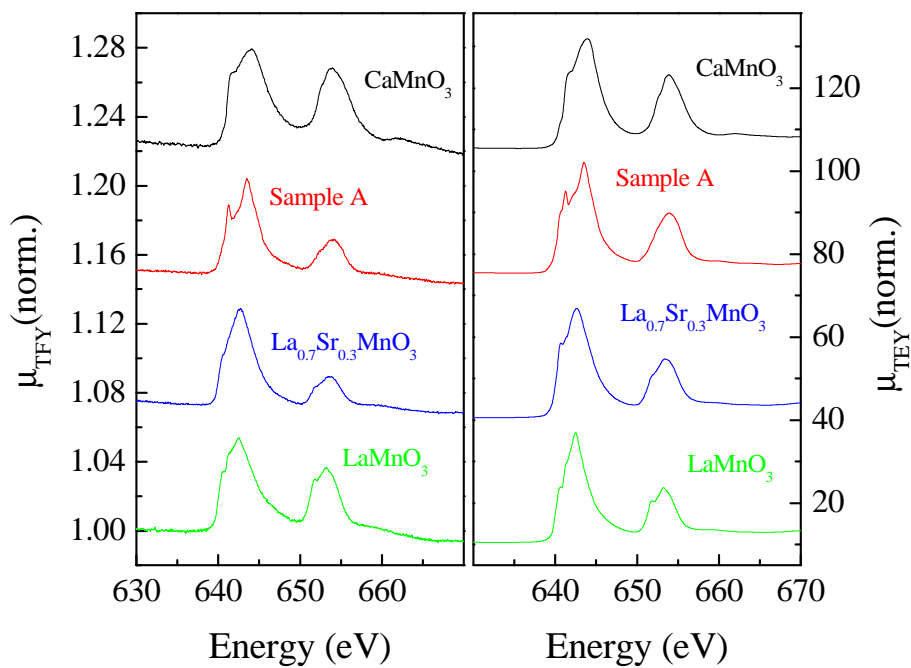

Figure 2. Absorption-corrected TFY-detected Mn $L_{2,3}$ XAS spectra of $CaMnO_3$, LCMO, LSMO and LMO at 300 K. Spectra have been vertically shifted for clarity.

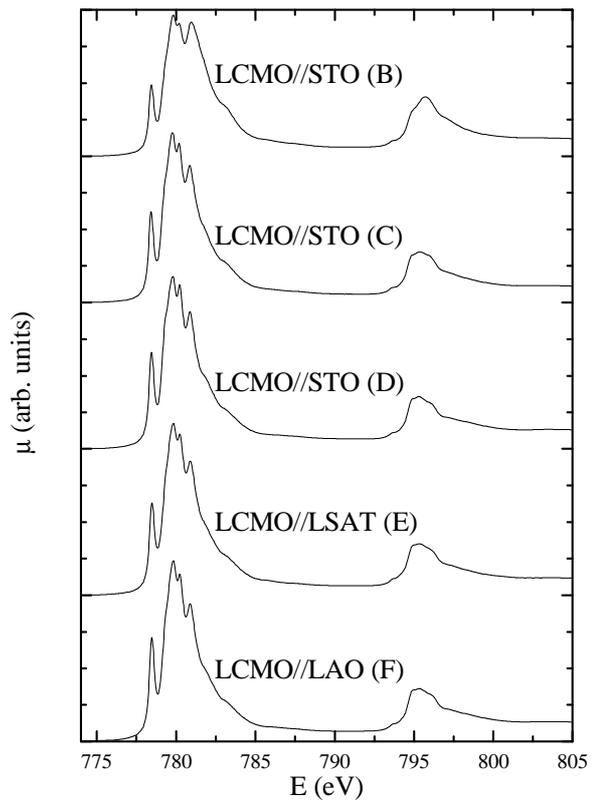

Figure 3:Co $L_{2,3}$ XAS spectra of Samples B to F as measured by TEY at 300 K. Spectra have been vertically shifted for clarity.

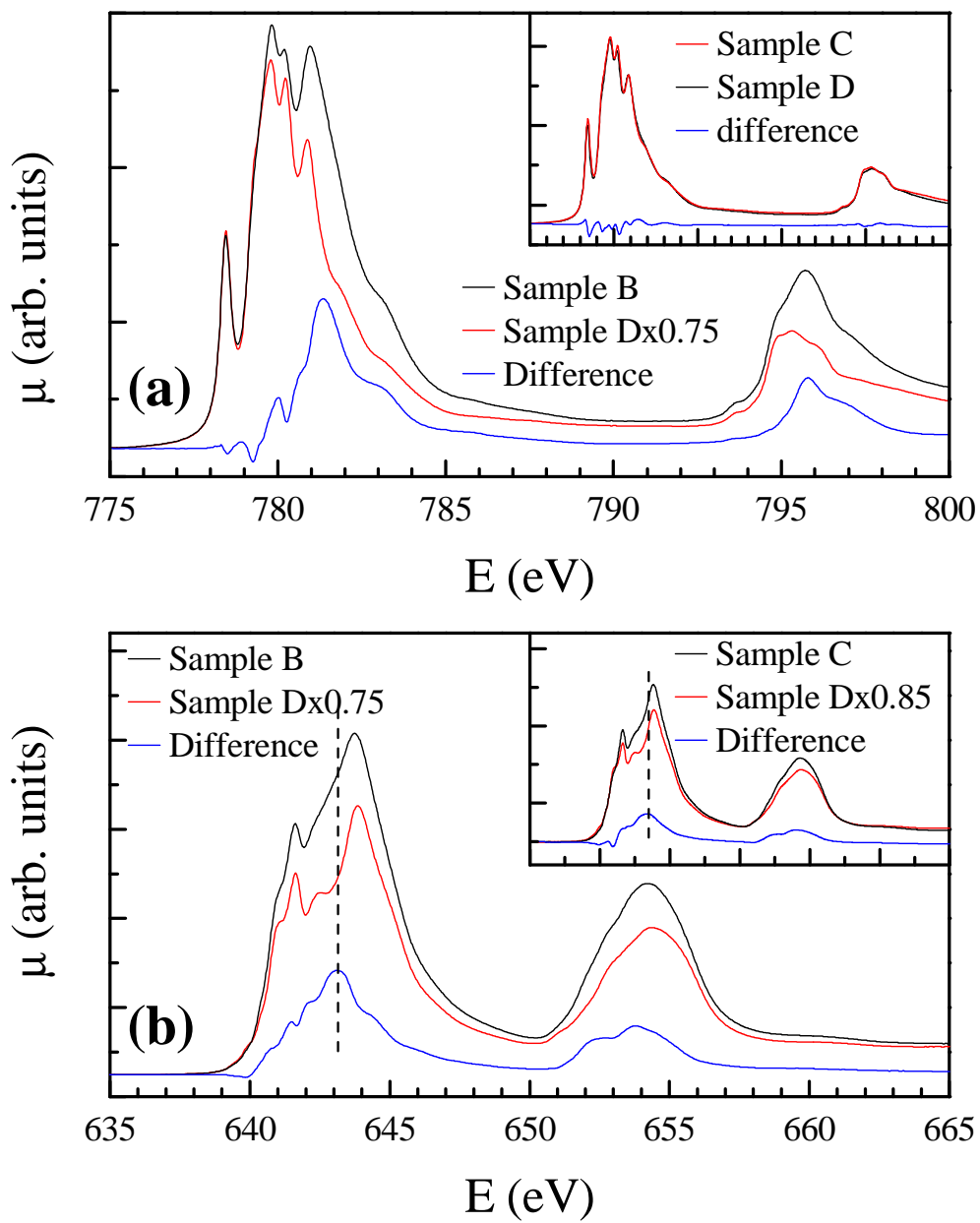

Figure 4: (a) Co-$L_{2,3}$ XAS of samples B and D (the last scaled by a factor 0.75) and the difference between them. The inset compares Co-$L_{2,3}$ spectra of samples C and D and plots its difference. (b) Mn-$L_{2,3}$ XAS of samples B and D (the last scaled by a factor 0.75) and the difference between these two lines. The inset compares Mn-$L_{2,3}$ spectra of samples C and D (this last scaled by a factor 0.85) and the difference between these two. All plotted spectra were collected at 300K.

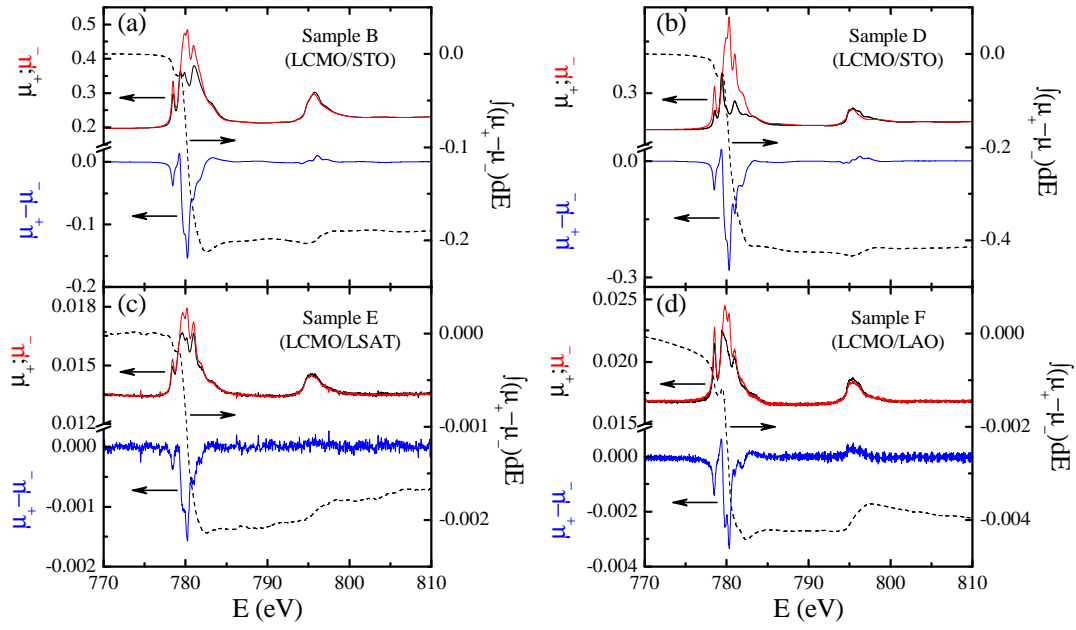

Figure 5: Co $L_{2,3}$ XMCD spectra of samples B and D (over STO) and samples E and F (over LSAT and LAO respectively) at T= 20K.

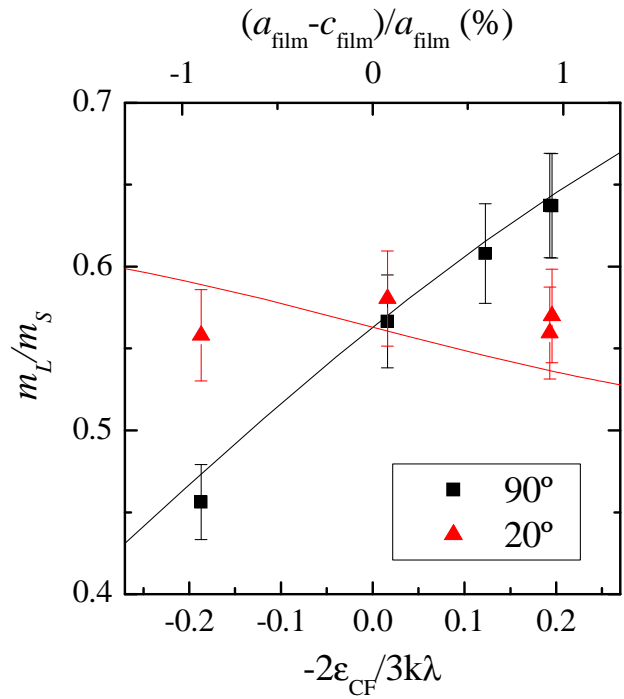

Figure 6: Comparison between measured $m_L/m_{Seff}$ ratios (symbols) as a function of $(a_F-c_F)/a_F$ (top x-axis) and predicted $m_L/m_S$ (solid lines with color according to symbols' one) as a function of $a(=\frac{2\epsilon_{CF}}{-3k\lambda})$ parameter (bottom x-axis). Both x-axes are in linear scale.